\input phyzzx.tex

\twelvepoint


\def\lambdabar{{\bar \lambda}}
\def\bz{{\bar z}}
\def\bs{{\bar s}}
\def\ba{{\bar a}}
\def\bu{{\bar u}}

\def\s{\hat s}


\REF\Wittenone{E. Witten, Nucl. Phys. {\bf B500} (1997) 3, hep-th/9703166}
\REF\HLW{P.S. Howe, N.D. Lambert and P.C. West, {\it Classical M-Fivebrane
Dynamics and Quantum $N=2$ Yang-Mills}, hep-th/9710034}
\REF\three{P.S. Howe, N.D. Lambert and P.C. West, {\it The Threebrane Soliton
of the M-fivebrane}, hep-th/9710033}
\REF\Wittentwo{E. Witten, {\it The Five-Brane Effective
Action in M theory}, hep-th/9610234}
\REF\Witten{E. Witten, private communication}
\REF\HSW{P.S. Howe, E. Sezgin and P.C. West, Phys. Lett. {\bf B399}
(1997) 49, hep-th/9702008}
\REF\HS{P.S. Howe and E. Sezgin, Phys. Lett. {\bf B394} (1997) 62,
hep-th/9611008}
\REF\BS{E. Bergshoeff and E. Sezgin, Nucl. Phys. {\bf B422} 
(1994) 329, hep-th/9312168}
\REF\S{M. Perry and J.H. Schwarz, Nucl. Phys. {\bf B489} (1997) 47,
hep-th/9611065; M. Aganagic, J. Park, C. Popescu, and J. H. Schwarz,
Nucl. Phys. {\bf B496} (1997) 191,
hep-th/9701166; P. Pasti, D. Sorokin and M. Tonin, Phys. Lett. {\bf B398}
(1997) 41, hep-th/9701037;
I. Bandos, K Lechner, A. Nurmagambetov, P. Pasti and D. Sorokin, and M.
Tonin, Phys. Rev. Lett. {\bf 78} (1997) 4332,
hep-th/9701149}
\REF\SW{N. Seiberg and E. Witten, Nucl. Phys. {\bf B426} (1994) 19, 
hep-th/9407087}
\REF\Grisaru{B. de Wit, M.T. Grisaru and 
M. Rocek, Phys. Lett. {\bf B374} (1996) 297, hep-th/9601115}
\REF\BSTY{A. Brandhuber, J. Sonnenschein, S. Theisen and 
S. Yankielowicz, Nucl. Phys. {\bf B504} (1997) 175, hep-th/9705232}
\REF\dHOO{J. de Boer, K. Hori, H. Ooguri and Y. Oz. {\it K\"ahler Potential
and Higher Derivative Terms from M theory Fivebrane}, hep-th/9711143}

\pubnum={KCL-TH-97-69  \cr hep-th/9712040}
\date{December 1997}

\titlepage

\title{\bf Gauge Fields and M-Fivebrane Dynamics}

\centerline{N.D. Lambert}

\centerline{and}

\centerline{P.C. West\foot{lambert, pwest@mth.kcl.ac.uk}}
\address{Department of Mathematics\break
         King's College, London\break
         England\break
         WC2R 2LS\break
         }

\abstract

In this paper we obtain both the vector and scalar equations of motion
of an M-fivebrane in the presence of threebrane solitons. The resulting
equations of motion are precisely those obtained from the Seiberg-Witten
low energy effective action for $N=2$ Yang-Mills, including all quantum
corrections. This analysis extends the work of a previous paper which
derived the scalar equations of motion but not in detail the vector
equations. We also discuss some
features of an infinite number of higher derivative terms predicted by 
M theory.

KEY WORDS: M Theory, Fivebrane, Yang-Mills

PACS: 11.15.Tk, 11.25.Sq, 11.30.Pb

\endpage


\chapter{Introduction}

Recently classical, long wavelength information of M theory has been 
successfully used to understand detailed quantum properties
of four-dimensional supersymmetric gauge theories. This work was initiated
by Witten [\Wittenone], who argued that the auxiliary Riemann surface
which appears in the Seiberg-Witten low energy effective action of
$N=2$ Yang-Mills can be naturally interpreted and moreover derived from
the geometric structure of a single M theory fivebrane. 

In a recent paper
it was shown [\HLW] 
that not only the geometrical structure of the Seiberg-Witten
solution, but in fact the entire low energy effective action can
be deduced from a knowledge of the classical  M-fivebranes equations of 
motion describing threebrane solitons [\three]. 
However, while  the full low energy
effective action was derived in [\HLW], the discussion  
relied heavily on the $N=2$ supersymmetry to deduce
the vector zero mode equations from a knowledge of the scalar action alone. 
In particular this  
calculation did not discuss  the origin of the vector zero
modes in detail  
or how their equations might be determined independently of the
scalar zero modes. Furthermore this calculation only made use of
the purely scalar part of the M-fivebrane equations, which can be derived
from the standard `brane' action $\sqrt{-\det g}$. Thus no use was made
of the rich and unique structure of the M-fivebrane arising from the
self-dual three-form. 
The purpose of this paper is to explicitly derive the
Seiberg-Witten effective action from the M-fivebranes equations of motion
for both the scalar and vector fields. 
In addition the literature has also been 
concerned with deriving information on $N=1$ Yang-Mills 
from the M-fivebrane. Another motivation for considering  the method 
presented in this paper is  
as a first step towards calculating the M-fivebrane's low energy
effective action in these cases.  It is also not known to
what extent the M-fivebrane can reproduce the correct low energy quantum
corrections to the classical action in these cases. 
Such calculations are then potentially particularly 
significant 
since there is no known analogue of the Seiberg-Witten effective action
for $N=1$ Yang-Mills coupled to massless scalars.

It is commonly written  that the vector part of the
Seiberg-Witten effective action may be obtained from  the
six dimensional action
$$
S = \int d^6 x\  H\wedge\star H\ ,
\eqn\thewrongthingtodo
$$
as follows.
First one decomposes the three form 
into $H = F^I\wedge\lambda_I$ where $\lambda_I$, $I=1,...,N$
are a basis of non-trivial one forms of the Riemann Surface and $F^{I}$ 
are four-dimensional $U(1)$ field strengths. The next step is to 
impose the self-duality constraint on $H$ and dimensionally 
reduce the resulting expression in \thewrongthingtodo\ over 
the Riemann surface $\Sigma$. However, 
as a text book calculation quickly reveals, since 
$H$ is an odd form $H\wedge \star H = H\wedge H = 0$. Thus it is highly
unlikely that an interesting action in four dimensions can be obtained by
this procedure. Another problem with an action as a starting point
can be seen in the case that the Riemann surface has genus one. One then  
has only one holomorphic form $\lambda$ and its complex conjugate $\lambdabar$.
Thus the only non-zero  expression that occurs is  
$\int_{\Sigma} \lambda\wedge\lambdabar$. However this integral  is
pure imaginary and is related to  ${\rm Im} \tau$ whereas both 
${\rm Im} \tau$ and ${\rm Re}\tau$ appear in the Seiberg-Witten 
effective action.

Indeed, there are strong objections to the use of an 
action for the M-fivebrane equations of motion, 
due to the chiral nature of its three form. In fact it has been shown that
no action can capture the full physics of the M-fivebrane 
[\Wittentwo]. This point has again been stressed more recently  by
Witten [\Witten].
In this paper we shall only consider the
equations of motion of the M-fivebrane and not attempt to
invoke an action at any time. At the end of the day it will turn out
that the resulting four-dimensional equations of motion do possess an
action formulation 
(the Seiberg-Witten effective action), but this is not
surprising since there are no longer any chiral forms.

In this paper  we shall use the manifestly covariant form of the 
M-fivebrane's equations of motion found in [\HSW]. These equations were derived
from the superembedding formalism applied to the M-fivebrane [\HS,\BS]. In the
next section we obtain the relevant equations for the low energy motion of
$N$ threebrane solitons moving in the M-fivebrane. In section three we discuss
the solution to the self-duality condition. In sections four and five 
we reduce the vector and scalar equations over the Riemann surface 
respectively. In section six we discuss the exact form of an infinite number
of higher derivative contributions to the purely 
scalar part of the low energy action, as predicted by the M-fivebrane. These
terms are compared with the corrections to the Seiberg-Witten effective
action obtained from calculations in Yang-Mills theory. Finally we close
with a discussion of our work in section seven.

It should be noted that there are other 
formulations of the M-fivebrane equations, which furthermore can be
derived from an action [\S]. This action 
relies on the appearance of an auxiliary, closed vector field $V$ with  unit 
length.\foot{There is a non-covariant form without an auxiliary
field, however this may be viewed a resulting from a particular choice of
the vector field, namely $V = dx^5$.} While the form of this action is
different from \thewrongthingtodo\ and so not manifestly zero, 
it is not clear what one should
take for the vector field $V$. Indeed on a generic Riemann Surface 
(with genus different from one) there are topological obstructions to the 
global existence of $V$. Once this problem is over come, one would
have to determine the r\^ole of the auxiliary field in relation to
the Seiberg-Witten effective action. As mentioned above, the 
derivation presented in this paper does not invoke a six-dimensional action.
Furthermore, some details of the derivation presented here could be 
interpreted  as suggesting that there are substantial difficulties 
in obtaining the Seiberg-Witten effective
action from a six-dimensional action. This maybe in accord with the statement
[\Wittentwo] that one cannot derive all of the M-fivebrane
physics from an action.


\chapter{Fivebrane Dynamics in the Presence of Threebranes}

The M theory fivebrane  has a six-dimensional $(2,0)$
tensor multiplet of
massless fields on its worldvolume. The component fields of this
supermultiplet are five real scalars $X^{a'}$, a gauge field $B_{\hat m \hat n}$
whose filed strength satisfies a modified self-duality condition and
sixteen spinors $\Theta ^i_\beta$. The scalars are the
coordinates  transverse to the fivebrane and correspond to the
breaking of 11 dimensional  translation invariance by the presence
of the fivebrane. The sixteen spinors correspond to the breaking
of half of the 32 component supersymmetry of M-theory.  The
classical equations of motion  of the fivebrane in the absence of
fermions and background fields are [\HSW]
$$
G^{\hat m\hat n} \nabla_{\hat m} \nabla_{\hat n} X^{a'}= 0\ ,
\eqn\eqomone
$$
and
$$
 G^{\hat m \hat n} \nabla_{\hat m}H_{\hat n\hat p\hat q}  = 0.
\eqn\eqomtwo
$$
where the worldvolume indices are $\hat m,\hat n, \hat p=0,1,...,5$
and  the world tangent indices $\hat a,\hat b,\hat c=0,1,...,5$.
The transverse indices are $a',b'=6,7,8,9,10$. We now define the
symbols that occur in the equation of motion. The usual induced metric
for a $p$-brane is given, in static gauge,  by
$$
g_{\hat m \hat n} = \eta _{\hat m \hat n}+ 
\partial _{\hat m}X^{a'} \partial _{\hat n}X^{b'}\delta _{a' b'}\ .
\eqn\gdef
$$
The covariant derivative in the equations of motion
is defined with the Levi-Civita connection with respect to the metric
$g_{\hat m \hat n} $.
Its action on a vector field $T_{\hat n}$ is given by
 $\nabla_{\hat m} T_{\hat n} = \partial _{\hat m} T_{\hat n}-
 \Gamma _{\hat m \hat n}^{\hat p}T_{\hat p}$
where
$$
\Gamma _{\hat m \hat n}^{\ \ \hat p}
= \partial _{\hat m } \partial _{\hat n} X^{a'}
\partial _{\hat r} X^{b'}g^{\hat r \hat s}\delta _{a' b' }\ .
\eqn\Gammadef
$$
We define the vielbein associated with the above metric in the
usual way
$g_{\hat m\hat n}=
e_{\hat m}^{\ \hat a} \eta _{\hat a \hat b} e_{\hat n}^{\ \hat b}$. 
The inverse metric $G^{\hat m\hat n}$ which occurs in the equations
of motion
is related to the usual induced metric given above by the equation
$$
G^{\hat m\hat n} = {(e^{-1})}^{\hat m}_{\ \hat c} \eta ^{\hat c \hat a}
m_{\hat a}^{\ \hat d} m_{\hat d} ^{\ \hat b} {(e^{-1})}^{\hat m}_{\  \hat b}\ .
\eqn\Gdef
$$
The matrix $m$ is given by
$$
m_{\hat a}^{\ \hat b} = \delta_{\hat a}^{\ \hat b}
 -2h_{\hat a\hat c\hat d}h^{\hat b\hat c\hat d}\ . 
\eqn\mdef
$$
The field $h_{\hat a\hat b\hat c}$ is an anti-symmetric three form
which is self-dual;
$$
h_{\hat a\hat b\hat c}=
{1\over3!}\varepsilon_{\hat a\hat b\hat c\hat d\hat
e\hat f}h^{\hat d\hat e\hat f}\ ,
\eqn\hsd
$$
but it is not the curl of a three form gauge field. It is related to
the field
$H_{\hat m \hat n \hat p}$ which appears in the
equations of motion
and is the curl of a gauge field,  but
$H_{\hat m \hat n \hat p}$ is not self-dual.
The relationship between
the two fields is given by
$$ 
H_{\hat m \hat n \hat p}= e_{\hat m}^{\ \hat a}
e_{\hat n}^{\ \hat b} e_{\hat p}^{\ \hat c} {({m }^{-1})}_{\hat
c}^{\ \hat d} h_{\hat a\hat b\hat d}\ .
\eqn\Hh
$$
Clearly, the self-duality condition on $h_{\hat a\hat b\hat d}$
transforms into a 
condition on $H_{\hat m \hat n \hat p}$ and vice-versa 
for the Bianchi identify $dH=0$. 

The appearance of a metric in the equations of motion which is
different to the usual induced metric has its origins in the
fact that the natural metric that appears for the fivebrane
has an associated  inverse vielbein  denoted by 
${(E^{-1})}_{\hat a}^{\ \hat m}$ which
is related in the usual way through $G^{\hat m\hat n} =
{(E^{-1})}_{\hat a}^{\ \hat m}
{(E^{-1})}_{\hat b}^{\ \hat n}\hat \eta^{\hat a\hat b}$.
The relationship between the two inverse vielbeins being
${(e^{-1})}_{\hat a}^{\ \hat m} =(m^{-1})_{\hat a}^{\ \hat b}
{(E^{-1})}_{\hat b}^{\ \hat
m}$. The inverse vielbein ${(E^{-1})}_{\hat a}^{\ \hat m}$ will play no 
further 
role in this paper. 
This completes our discussion of the fivebrane equations of
motion and we refer the reader to reference [\HSW]  for more details
of the formalism and notation.

We will be interested in fivebrane configurations that contain within 
them threebrane solutions. Such solutions, which were  found in [\three], 
play an crucial  part in the recovery of the
$N=2$ Yang-Mills theory in four dimensions and   we now
summarise them. Given that the   six-dimensional
coordinates of the fivebrane are denoted by
hatted variables $\hat m,\hat n = 0,1,2,...,5$
we take the world-volume
of the threebrane to be   in the plane $x^\mu=(x^0,x^1,x^2,x^3)$. We
let  unhatted variables refer to the coordinates transverse
of the threebrane i.e. $x^n=(x^4,x^5)$. We will assume all
fields to depend only on these transverse coordinates. In fact, of the
two transverse scalars of the fivebrane we take only two of them,
$X^6$ and $X^{10}$ to be non-constant.
We also take
the gauge field strength
$H_{\hat \mu \hat \nu \hat \rho}=0$. Examining the
supersymmetric variation of the spinor we find that this configuration
preserves half of the original sixteen supersymmetries, leaving eight
supersymmetries,  provided the two transverse coordinates $X^6$ and
$X^{10}$ obey the Cauchy-Riemann equation with respect to $x^4$ and
$x^5$ [\three]. As such, we introduce the variables
$$ 
z= x^4 + i x^5,\ \ s= {X^6+iX^{10}} \ ,
\eqn\zsdef
$$
and conclude that  the Bogomoln`yi condition is simply that
$s$ depends only on $z$ and not $\bar z$ (i.e.
$s=s(z)$).  As seen from the M-theory perspective this result means
that  the presence of threebranes within the the fivebrane 
implies that the fivebrane is
wrapped  on a Riemann surface in the space with complex coordinates
$s$ and $z$ [\HLW]. In fact, the threebranes correspond to the self
intersections of the M-theory fivebrane.

The fivebrane equations then reduce to the
flat Laplacian [\three]
$$
\delta^{mn} \partial _n \partial _m s =0\ ,
\eqn\flatlap
$$
which is automatically satisfied due to the Bogomoln'yi condition.
We are interested in field configurations which are everywhere
smooth except at $z=\infty$.  Some such configurations are  given
by [\HLW]
$$
s = s_0-\ln\left(B +\sqrt{Q} \right)\ ,
\eqn\s
$$
where we  have introduced the polynomial
$Q = B^2(z) - \Lambda^{2N}$ and
$s_0$ and $\Lambda$
are constants.
The quantity  $B(z)$ is a $N$th order polynomial in $z$ which can be
written in the form
$$
B(z) = z^{N} - u_{N-1} z^{N-2} - u_{N-2} z^{N-3} - ... - u_{1}\ .
\eqn\Bdef
$$
Following [\Wittenone] we introduce the variable $t= e^{-s}$ 
whereupon we find
that the threebrane solution implies the equation
$$
F(t,z) = t^2 - 2B(z)t + \Lambda^{2N} = 0 \ .
\eqn\Ftwo
$$
We recognise this equation, after a
suitable shift in $t$ to absorb the term linear in t, to be
the standard equation for a
hyper-elliptic curve of genus
$N$. The
$u_i$'s corresponding to the moduli of this Riemann surface.
For simplicity, we will be interested for most of  this paper in
the case of $N=2$, but the results are readily extended to
the case of $N\ge 3$

We are interested in the low energy motion of  $N$ threebranes and so
take the zero modes of the threebrane to depend on
the  worldvolume coordinates $x^\mu,\ \mu=0,1,\dots 3$, of the threebrane. 
Thus we will arrive at a theory with eight supersymmetries  
living on the four dimensional threebrane 
worldvolume. 
The moduli $u_i$ of the Riemann surface are related to the positions of
the $N$ threebranes  and as such we take them to depend on the
world-volume  i.e. 
$u_i(x^{\mu})$. For example the $u_N$ coefficient, which is the sum of the
roots of $B$, represents the centre of mass coordinate for the threebrane and
has been set to zero. 
These will  turn out to be the $2N$ complex scalars in the four
dimensional theory.  There are also
$4N$ further bosonic moduli associated with large two form gauge
transformations at infinity. These are associated with the field
strength
$H_{\hat m\hat n\hat p}$ and their precise form will  be discussed
later. These zero modes   correspond to the gauge fields of
the resulting four dimensional theory on the four
dimensional threebrane world-volume.  We also have $8N$ fermionic  zero modes
corresponding to the broken supersymmetries in the presence of
the threebrane, these correspond to the spinors in the resulting four
dimensional theory. Taken together the zero modes  make up
a $N=2$ super ${(U(1))}^N$ multiplet which lives on  the four dimensional
threebrane world-volume.

This procedure is analogous to the more simple case of
monopole solutions to
$N=2$ Yang-Mills theory in four dimensions.  At low energy the
 motion of the monopoles can be described by taking the
moduli of the monopole solution to depend on the worldline
coordinate $t$. For the case of one monopole the moduli
space is just given by the transverse coordinates
$\underline x$ and the coordinate corresponding to large gauge
transformations $\theta$.

Our main task in this section is to derive the consequences of the
fivebrane classical dynamics, encoded in the equations of motion
\eqomone\ and \eqomtwo , 
for the $x^\mu$ dependent moduli of the threebrane.
The first step
in this direction is to work out in detail the geometry of the
fivebrane  in the presence of these zero modes. We  work with the
coordinates $z = x^4+ix^5\quad\bar z=x^4-ix^5$, 
introduced above,  for
which the Euclidean  metric takes the form
$\eta_{z\bar z}={1\over2}\quad\eta^{z\bar z}=2$ and
$\eta_{z z}=0 =\eta_{\bar z\bar z}$. We also define the derivatives
$\partial= {\partial \over \partial z}$ and
$\bar \partial= {\partial \over \partial \bar z}$.  In these
coordinates we find, for example,  that
$ |\partial s|^2=\partial s\bar \partial \bar s
={1\over2}\delta ^{mn}\partial_ns\partial_m\bar s$.

The usual induced metric of the fivebrane, in the static gauge,  and
in the presence of the threebrane  takes the form
$$
g_{\hat n\hat m}
=\eta_{\hat n\hat m}+{1\over2}(\partial_{\hat
m}s\partial_{\hat n}\bar s+\partial_{\hat n}s\partial_{\hat m}\bar s)\ .
\eqn\gbrane
$$
For future use we list the individual components in the
longitudinal and transverse directions to the threebrane
$$\eqalign{
g_{z\mu}&={1\over2}\partial s\partial_\mu\bar s= {(g_{\bar z \mu})}^*\ ,\cr
g_{\mu \nu}&=\eta_{\mu\nu}+{1\over2}(\partial_\mu
s\partial_\nu\bar s+\partial_\nu s\partial_\mu\bar s)\ ,\cr
g_{z\bar z}&= g_{\bar z z}={1\over 2}(1+|\partial s|^2)\ ,\cr
g_{\bar z\bar z}&= g_{z z} = 0\ .\cr}
\eqn\gcomplex
$$
It is straightforward, if a little tedious,  to construct the
inverse of this metric and  the result is
$$
g^{\hat m\hat n}
=\eta^{\hat m\hat n}+\alpha(\partial^{\hat
m}s\partial^{\hat n}\bar s
+\partial^{\hat m}\bar s\partial^{\hat n}s)
+\beta\partial^{\hat m}s\partial^{\hat n}s
+\bar\beta\partial^{\hat
m}\bar s\partial^{\hat n}\bar s\ ,
\eqn\ginv
$$
where
$$\eqalign{
\alpha&={1\over2}{(1+|\partial s|^2+{1\over2}|\partial_\mu
s|^2)\over\left\{{1\over4}(\partial_\mu\bar s)^2(\partial_\mu
s)^2-\big(1+|\partial s|^2+{1\over2}(|\partial_\mu
s|)^2\big)^2\right\}} \ ,\cr
\beta
&=-{\partial_\mu\bar s \partial^\mu\bar s\over\left\{(\partial_\mu\bar
s)^2(\partial_\mu s)^2-4(1+|\partial s|^2+{1\over2}|\partial_\mu
s|^2)^2\right\}}\ .\cr}
\eqn\alphabetadef
$$
We will only be interested in the low energy action and 
it will prove  useful to list the
component of the inverse metric to the second order in spacetime derivatives. 
The results are given by
$$\eqalign{
g^{\mu\nu}&=\eta^{\mu\nu}
+{1\over2}(1+|\partial s|^2)^{-1}(\partial^\mu s\partial^\nu\bar s
+\partial^\mu\bar s\partial^\nu s)
+O((\partial_\mu s)^3)\ , \cr
g^{\mu z}&= -(1+|\partial s|^2)^{-1}\partial _\mu s \partial \bar
s +O((\partial_\mu s)^3) = {(g^{\mu \bar z})}^*\ ,\cr
g^{zz}&= (1+|\partial s|^2)^{-2} \partial _\rho  s
 \partial ^\rho s
\bar \partial \bar s\bar  \partial\bar  s +
O((\partial_\mu s)^3)=  {(g^{\bar z \bar z})}^*\ , \cr  
g^{z\bar z}&={2 \over 1 + |\partial s|^2} 
+ {\partial_{\mu}s\partial^{\mu}\bs|\partial s|^2\over (1 + |\partial s|^2)^2}
+ O((\partial_\mu s)^3)\ .\cr}
\eqn\glowest
$$

We also require the vielbein associated with the usual induced
metric (i.e.
$e_{\hat n}{}^{\hat a}\eta_{\hat a\hat b}e_{\hat m}{}^{\hat b}
=g_{\hat n\hat m}$).
To the order in spacetime derivatives to which we are working
the components of the vielbein is given by
$$\eqalign{
e_\mu{}^z
&={1\over(1+|\partial s|^2)^{1\over2}}\partial_\mu s
{\bar \partial} \bar s;\quad e_z{}^\mu=0 \ ,\cr
e_{\mu}^{\bar z}&={1\over(1+|\partial s|^2)^{1\over2}}\partial_\mu\bar
s\partial s;
\quad e_{\bar z}^{\ \mu}=0 \ ,\cr
e_z ^{\ z} &= e_{\bar z}^{\ \bar z} = (1+|\partial s|^2)^{1\over2} \ ,\cr
e_\mu ^{\ a}&=\delta_\mu ^{\ a}\ ,\quad 
e_{\bar z}^{\ z}=e_{z}^{\ \bar z}=0 \ .\cr}
\eqn\vielbein
$$
Finally, we compute the Christoffel symbol given in equation \Gammadef. For
the configuration of interest to us  it becomes
$$
\Gamma_{\hat n\hat m}^{\ \ \ \hat r}
=(\partial_{\hat n}\partial_{\hat m}
X^{a'})\partial_{\hat p}
X_{a'}g^{\hat p\hat r}
={1\over2}(\partial_{\hat n}\partial_{\hat m} s
\partial_{\hat p}\bar s g^{\hat p\hat r}+\partial_{\hat
n}\partial_{\hat m}\bar s\partial_{\hat p} s g^{\hat p\hat r})\ .
\eqn\Gammais
$$
To the order to which we are working we find, for example,  that
$$\eqalign{
\Gamma_{\mu\nu}{}^{\rho}&=0\ ,\cr
\Gamma_{\mu\nu}{}^z&
=\partial_{\mu}\partial_{\nu} s\bar \partial \bar s
(1+|\partial s|^2)^{-1}\ ,\cr
\Gamma_{\mu\nu}{}^{\bar z}
&=\partial_\mu\partial_\nu\bar s\partial s (1+|\partial s|^2)^{-1}\ .\cr}
\eqn\Gammalowest
$$

Having computed the geometry of the fivebrane in the presence of the
threebrane zero modes, we can now evaluate the bosonic equations of
motion for the fivebrane. While the order of the the $\partial _\mu s$ 
is clear form the above expressions we must also establish the order 
of the spacetime derivatives in  the 
gauge field field strength $H_{}$. The precise form of this 
object follows from solving the self-duality condition and is 
given in the next section. We note here  that 
$H_{\mu\nu z}= {(H_{\mu\nu \bar z}})^*$ is first 
order in spacetime derivatives, while $H_{\mu\nu\rho}$ is  second order
in spacetime derivatives and $H_{\mu z\bz}=0$. 

We begin with the scalar equation
of equation \eqomone. Using \mdef\ to second order in spacetime 
derivatives this equation can be written as 
$$
g^{\hat m\hat n}\nabla_{\hat m}\nabla_{\hat n} s
-4(e^{-1})^{z}_{\ z}h^{z\hat c\hat d}h_{z\hat c\hat d}
(e^{-1})^z_{\ z}\nabla_z \nabla_z s =0\ .
\eqn\scalarone
$$
Using equation \Gammalowest\ for the 
Christoffel symbol the
term $g^{\mu \nu}\nabla_\mu \nabla_\nu s$  becomes
$$
g^{\mu\nu}\nabla_\mu\nabla_\nu s
=g^{\mu\nu}(\partial_\mu\partial_\nu
s-\Gamma_{\mu\nu}{}^{\hat n}\partial_{\hat n}s)
=g ^{\mu\nu}\partial_\mu\partial_\nu s 
\left( 1- 
{1\over 2} \partial _{\hat p }\bar s g^{\hat p \hat n} \partial _{\hat n } s
\right) \ ,
\eqn\scalartwo
$$
The final factor in this equation can be evaluated 
to be ${(1+|\partial s|^2)}^{-1}$. The other terms 
can be processed in a similar way and one finds that the scalar 
equation of motion is given by
$$
(g^{\mu\nu}\partial_\mu\partial_\nu s+2g^{\mu
z}\partial_\mu\partial_zs+g^{zz}\partial_z\partial_zs
-4(e^{-1})^{z}_{\ z}h^{z\hat c\hat d}h_{z\hat c\hat d}
(e^{-1})^z{}_{\ z}\partial \partial s) 
{1\over(1+|\partial s|^2)}=0\ .
\eqn\scalarthree
$$

Substituting the expressions for the 
inverse metric to the appropriate order in (four-dimensional) 
spacetime derivatives, we
find  the scalar equation becomes
$$
{1\over(1+|\partial s|^2)} E=0\ , 
\eqn\scalarfour
$$
where
$$
E\equiv \eta^{\mu\nu}\partial_\mu\partial_\nu s
-\partial_z\left\{{(\partial_\varrho s\partial^\varrho s)
\bar \partial\bar s\over(1+|\partial s|^2)}\right\}
-{16\over(1+|\partial s|^2)^2}
H_{\mu\nu\bar z}H^{\mu\nu}{}_{\bar z}\partial \partial s
=0\ .
\eqn\Escalar
$$

Let us now evaluate the vector equation \eqomtwo. To the order in
spacetime derivatives to which we are working we can set
$m^{\ \hat a}_{\hat b}= \delta  ^{\ \hat a}_{\hat b}$
and the vector equation becomes
$$
g^{\hat m\hat n}\nabla_{\hat m}H_{\hat n\hat p\hat q}=0\ .
\eqn\vectorone
$$
Taking $(\hat p,\hat q)=(\mu,\nu)$
the equation can be shown to become
$$E_{\mu\nu}\equiv
\partial_zH_{\mu\nu\bar z}+\partial_{\bar z}H_{\mu\nu z}=0\ ,
\eqn\Eoff
$$
after discarding a the factor $(1+|\partial s|^2)^{-1}$. 
In finding this last result we 
have, for example, discarded spacetime derivatives acting on 
$H_{\mu\nu\rho}$
as such terms would be cubic in spacetime derivatives. 

Taking  $(\hat p,\hat q)=(\nu,z)$ the equation becomes
$$
g^{\hat m \hat n}\{\partial_{\hat m}H_{\hat n\nu z}-\Gamma_{\hat
m \hat n}{}^{\hat p}
H_{\hat p\nu z}-\Gamma_{\hat m\nu}{}^{\hat p}H_{\hat n
pz}-\Gamma_{\hat m z}{}^{\hat p}H_{\hat n \nu\hat p}\}=0\ .
\eqn\vectortwo
$$
These terms can be processed as for the scalar equation,
for example, we find that
$$ 
-g^{\hat m \hat n }\Gamma_{\hat m z}^{\ \ \ \hat r}H_{\hat n \nu \hat r}
=- \partial_z \left(
{\partial^\mu s\bar \partial \bar s\over
(1+|\partial s|^2)}\right) H_{\mu\nu z}
+ \partial_z \left(
{\partial^\mu \bs \partial s\over
(1+|\partial s|^2)}\right) H_{\mu\nu \bz}\ .
\eqn\vectorfive
$$
Evaluating the other terms in a similar way the vector equation
becomes
$$
\partial^\mu H_{\mu\nu z}-{\partial_\mu\bar s
\partial s\over(1+|\partial s|^2)}\partial_{\bar z}H_{\mu\nu z}
-\partial_z\left\{{\bar \partial \bar s\partial_\mu sH_{\mu\nu z}
\over(1+|\partial s|^2)}\right\}
+\partial_z\left\{{\partial s\partial_\mu{\bar s}\over(1+|\partial
s|^2)}\right\}H_{\mu\nu\bar z}=0\ .
\eqn\vectorsix
$$
Finally, using equation \Eoff\ 
we can rewrite this as
$$
E_{\nu z}\equiv
\partial^\mu H_{\mu\nu z}-\partial_zT_{\nu}=0\ ,
\eqn\Evect
$$
where 
$$
T_{\nu} ={\bar \partial \bar
s\partial^\mu s\over(1+|\partial s|^2)}H_{\mu\nu  z}
-{\partial s\partial^\mu\bar s\over(1+|\partial s|^2)}H_{\mu\nu\bar z}\ .
\eqn\Tdef
$$


\chapter{The Self-Dual Three Form}

Although the field $H_{\hat m \hat n \hat p}$ is a curl it does not obey a 
simple self-duality condition. On the other hand,  
the field $h_{\hat a \hat b \hat c}$ 
is not a curl but does obey a simple self-duality condition,  
namely 
$$
h_{\hat a\hat b\hat c}={1\over3!}\varepsilon_{\hat a\hat b\hat c\hat d\hat
e\hat f}h^{\hat d\hat e\hat f}\ .
\eqn\hsd
$$
The strategy we adopt to solve the self-duality condition  for 
$h_{\hat a\hat b\hat c}$ and use equation \Hh\ which relates 
$H_{\hat m \hat n \hat p}$ to $h_{\hat a\hat b\hat c}$ and deduce 
the consequences for $H_{\hat m \hat n \hat p}$. 

We adopt the convention that all indices on $h_{\dots}$ are always tangent 
indices.
Upon taking the various choices for the indices ${\hat a\hat b\hat c}$ 
we find that equation \hsd\ becomes 
$$\eqalign{
h_{abz}&={i\over2}\varepsilon_{abcd}h^{cd}_{\ \ z}\ ,\cr
h_{ab\bar z}&=-{i\over2}\varepsilon_{abcd}h^{cd}_{\ \ \bar z}\ ,\cr
 h_{z\bar za}&={1\over3!}{i\over2}\varepsilon_{abcd}h^{bcd}\ ,\cr
 h_{bcd}&=-2i\varepsilon_{bcde}h_{z\bar z}^{\ \ \ e}\ .\cr}
\eqn\hsdcmpt
$$
Hence the independent components can be taken to be 
$h_{abz},\ h_{ab\bar z}$ and $h_{z\bar z e}$. 
We are not interested in the most general such three form,  but one which 
corresponds to the zero modes of the three form field 
strength that correspond to 
finite gauge transformations at infinity. As such, we 
set $h^{}_{z\bar ze}=h^{}_{abc}=0$. 

We now deduce $H_{\hat n\hat m\hat p}$ from
Equation \Hh\ which at the order  we are working takes the form 
$$
H_{\hat n\hat m\hat p}=e_{\hat n}{}^{\hat a}e_{\hat m}{}^{\hat b}
e_{\hat p}{}^{\hat c}h_{\hat a\hat b\hat c}\ .
\eqn\Hhtwo
$$
Taking $(\hat n\hat m\hat p)=(\mu,\nu,z)$, using the form of the  vielbein 
of equation \vielbein, and working to at most second order in 
spacetime derivatives we find that 
$$ 
H_{\mu\nu z}=\delta_\mu{}^a\delta_\nu{}^b(1+|\partial
s|^2)^{1\over2}h^{}_{abz}\ ,
\eqn\Hhthree
$$
since the vielbein  $e_z{}^{\hat c}$ can only have $\hat c=z$ to order   
$(\partial_\mu s)^1$. 
Taking $(\hat n\hat m\hat p)=(\mu,z,\bar z)$ equation \Hh\ becomes 
$$
H_{\mu z\bar z}= e_\mu{}^ae_z{}^{z}e_{\bar z}{}^{\bar z}
h^{}_{az\bar z}=0 \ .
\eqn\Hhfour
$$
Finally taking $(\hat n\hat m\hat p)=(\mu,\nu, \rho)$, we find that 
$$\eqalign{
H_{\mu\nu\varrho}&= 
3\delta_{[\mu}{}^a\delta_\nu{}^b(e_{\varrho]}{}^z
h_{abz}+e_{\varrho]}{}^{\bar z}h_{ab\bar z})\ , \cr
&={3\over (1+|\partial s|^2)^{1\over2}}
\delta_{[\mu}{}^a\delta_\nu{}^b(\partial_{\varrho]}s\partial_{\bar z}
{\bar s}h_{abz}+\partial_{\varrho]}\bar s\partial_zsh_{ab\bar z})\ .\cr}
\eqn\Hhfive
$$
Since $h_{abz}$ and $h_{abz}$ obey the self-duality conditions of equation 
\hsd, 
it follows that $H_{\mu\nu z}$ and $H_{\mu\nu \bar z}$ will obey the 
self-duality relations 
$$ 
H_{\mu\nu z}={i\over2}\varepsilon_{\mu\nu\varrho\kappa}
H^{\varrho \kappa}_{\ \ \  z};\  
H_{\mu\nu\bar z}=-{i\over2}\varepsilon_{\mu\nu\varrho \kappa}
H^{\varrho \kappa}_{\ \ \ \bar z}\ .
\eqn\Hsd
$$
Substituting for $H_{\mu\nu z}$ and $H_{\mu\nu \bar z}$ in equation \Hhfive\
we find that $H_{\mu\nu\rho}$ can be written as 
$$
H_{\mu\nu\varrho}={3\over(1+|\partial
s|^2)}(\partial_{[\varrho}s\bar \partial \bar
sH_{\mu\nu]z}+\partial_{[\varrho}\bar s\partial s H_{\mu\nu]\bar z})\ ,
\eqn\Hsdtwo
$$
which in turn can be rewritten as 
$$
H_{\mu\nu\varrho}={i\over(1+|\partial
s|^2)}\varepsilon_{\mu\nu\varrho\lambda}(\partial^\tau\bar
s\partial s H_{\lambda\tau\bar z}-\partial^\tau s\bar \partial {\bar s}
H_{\lambda\tau z})\ .
\eqn\Hoff
$$
In the previous section we worked out the equations of motion 
for the three form and in this section we have solved the self-duality 
condition arising in the fivebrane dynamics to find 
$$\eqalign{
H = & {1\over 2!}{H}_{\mu\nu z}dx^{\mu}\wedge dx^{\nu} \wedge dz
+ {1\over 2!}{\bar H}_{\mu\nu\bz}dx^{\mu}\wedge dx^{\nu} \wedge d\bz \cr
&+{1\over 2!}{i\over(1+|\partial s|^2)}
\varepsilon_{\mu\nu\varrho\lambda}(\partial^\tau\bar
s\partial_zsH_{\lambda\tau\bar z}-\partial^\tau s\partial_{\bar z}{\bar s}
H_{\lambda\tau z})
dx^{\mu}\wedge dx^{\nu} \wedge dx^{\varrho}\ .}
\eqn\Hform
$$
Finally we should
check that the three form 
$H$ is closed (to second order in four-dimensional derivatives): 
$$\eqalign{
dH &= 
{1\over 2!}\partial_{\lambda}{H}_{\mu\nu z}
dx^{\lambda}\wedge dx^{\mu}\wedge dx^{\nu} \wedge dz
+ {1\over 2!}\partial_{\lambda}{\bar H}_{\mu\nu\bz}
dx^{\lambda}\wedge dx^{\mu}\wedge dx^{\nu} \wedge d\bz\  \cr
&-{1\over 3!}\partial_{z}H_{\mu\nu\lambda}
dx^{\mu}\wedge dx^{\nu} \wedge dx^{\lambda}\wedge dz
-{1\over 3!}\partial_{\bz}H_{\mu\nu\lambda}
dx^{\mu}\wedge dx^{\nu} \wedge dx^{\lambda}\wedge d\bz \cr
&-{1\over 2!}\partial_{z}H_{\mu\nu\bz}
dx^{\mu}\wedge dx^{\nu} \wedge d\bz\wedge dz
+{1\over 2!}\partial_{\bz}H_{\mu\nu z}
dx^{\mu}\wedge dx^{\nu} \wedge dz\wedge d\bz\ ,\cr
&= {i\over 3!}\varepsilon^{\ \ \ \ \mu}_{\rho\lambda\nu}E_{\mu z}
dx^{\rho}\wedge dx^{\lambda}\wedge dx^{\nu} \wedge dz
- {i\over 3!}\varepsilon^{\ \ \ \ \mu}_{\rho\lambda\nu}E_{\mu \bz}
dx^{\rho}\wedge dx^{\lambda}\wedge dx^{\nu} \wedge d\bz \ ,\cr
&+ {1\over 2!}
 E_{\mu\nu}dx^{\mu}\wedge dx^{\nu}\wedge dz\wedge d\bz\ .\cr}
\eqn\Hclosed
$$
Thus 
one can  readily verify that conditions following from  $dH=0$ are just 
equations of motion found in the previous section, as indeed  
should be the case. 

We complete this discussion  by solving equation \Hsd\ in terms of a 
real field $F_{\mu\nu}$ and its four-dimensional 
dual $\star F_{\mu\nu}= {1\over 2}
\epsilon _{\mu\nu\lambda\tau}F^{\lambda \tau}$. Writing $H_{\mu\nu z}$ 
as an arbitrary linear combination of these two fields we find that the 
only solution to equation \Hsd\ is given by 
$$
H_{\mu\nu z} = \kappa {\cal F}_{\mu\nu}
\eqn\Hdef
$$
where ${\cal F}_{\mu\nu}= F_{\mu\nu} + i\star F_{\mu\nu}$ and 
$\kappa$ is an as yet undetermined quantity. 
Since $H_{\mu\nu \bar z}$ is the complex conjugate of 
$H_{\mu\nu z}$ we conclude that 
$$ 
H_{\mu\nu \bar z} = {\bar \kappa}{\bar {\cal F}}_{\mu\nu} 
\eqn\Hbardef
$$
where ${\bar {\cal F}}_{\mu\nu}= F_{\mu\nu} -i\star F_{\mu\nu}$.

In order to satisfy the equation \Eoff\ one a finds
that $\kappa dz$ must be a holomorhic one form on $\Sigma$. 
Therefore, for a Riemann surface of
genus one, we must set 
$\kappa = \kappa_0\lambda_z$, where
$\lambda _z = ds/du$
is the unique (up to scaling) 
holomorphic one form on $\Sigma$  
and  $\kappa_0$ is independent of $z$ and $\bz$.
In fact, we will take 
$$
H_{\mu\nu z} = {ds\over da}{\cal F}_{\mu\nu}
= \left({da\over du}\right)^{-1}  {\cal F_{\mu\nu}} \lambda_z \ .
\eqn\H
$$
Here $a$ is the scalar mode used in the Seiberg-Witten theory [\SW] 
which we will define below. We will also see below  that  $\kappa dz$ is a 
holomorphic one form whose integral around the $A$-cycle is normalised to one. 
Of course until one specifies $F_{\mu\nu}$ the coefficient $\kappa_0$  
has no independent meaning. 
However, with this choice it will turn out that the equations of motion 
imply that $F_{\mu\nu}$ satisfies a simple Bianchi identity and so 
can be written as the curl of the four dimensional gauge field.
We point out  that our final result for $H$ is significantly different 
to those  proposed in a number of recent works.


\chapter{The Vector Equation of Motion}

In this section we will obtain the four-dimensional equations of motion for
the vector zero modes of the threebrane soliton. 
It is important to assume that
$X^{10}$ is compactified on a circle of radius $R$ and redefine
$s =(X^6+iX^{10})/R$. This allows
the connection between the M-fivebrane and 
perturbative type IIA string theory and quantum 
Yang-Mills theory to be made for small $R$ [\Wittenone] (it also ensures
that $s(t)$ is well defined).
Also, we promote 
$\Lambda$ and $z$ to have dimensions of
mass which  facilitates a more immediate contact with
the Seiberg-Witten solution. Thus in using the previous formulae 
we must rescale
$s\rightarrow R s$ and $z\rightarrow \Lambda^{-2} z$ (except that 
$\lambda_z = ds/du$ as before).

We have seen in equation \Evect\ that the three form equation of motion
to lowest order is $E_{\mu z} dz-{\bar E}_{\mu \bz}d\bz=0$. 
To obtain the equation of motion for the 
vector zero modes in four dimensions it is instructive to perform the
reduction over the Riemann Surface in two ways. First consider the integral
$$
\eqalign{
0 & = \int_B \star dH \int_A \lambdabar 
- \int_A \star dH\int_B \lambdabar \ ,\cr
&=\int_B (E_{\mu z}dz-{\bar E}_{\mu \bz}d\bz)\int_A \lambdabar 
- \int_A (E_{\mu z}dz-{\bar E}_{\mu\bz}d\bz)\int_B \lambdabar \ ,\cr}
\eqn\bilinear
$$
here $A$ and $B$ are a basis of cycles of the Riemann surface. 

Before proceeding with the integrals in \bilinear\ it is necessary to
remind the reader of the scalar fields $a$ and $a_D$. In [\SW] it was shown
that a global description of the moduli space was given by a pair of local
coordinates $a(u)$ and $a_D(u)$ defined as the periods of a single holomorphic
form $\lambda_{SW}$ 
$$
a \equiv \int_{A} \lambda_{SW}\ , \ \ \ a_D \equiv \int_{B} \lambda_{SW}\ .
\eqn\aad
$$
Furthermore
the Seiberg-Witten differential $\lambda_{SW}$ is itself defined so that
$\partial \lambda_{SW}/\partial u$ $=\lambda$. From \s\ one can check that
$\lambda_z = \partial s/\partial u$.
From these definitions one sees that
$$
\tau = {\int_{B}\lambda\over\int_{A} \lambda} = {da_D/du\over da/du} \ .
\eqn\tdef
$$
Here one sees that the coefficient in \H\ serves to normalise the 
$A$-period of $\kappa dz$ to unity.

Returning to the equations of motion, if 
we substitute $H_{\mu\nu z}= R(ds/da){\cal F}_{\mu\nu}$ 
into  $E_{\mu z}$ and  \bilinear\ we find that the
terms involving $T_{\mu}$ combine to form a total derivative which can
be ignored in the line integrals.  The remaining terms can then  simply be 
evaluated to give
$$\eqalign{
0= &\partial^{\nu}{\cal F}_{\mu\nu}\left({da\over du}\right)^{-1}
\left({da\over du}{d\ba_D\over d\bu} - {da_D\over du}{d\ba\over d\bu}\right)\cr
&- {\cal F}_{\mu\nu}\partial^{\nu}u {d^2 a\over du^2}
\left({da\over du}\right)^{-2}
\left({da\over du}{d\ba_D\over d\bu} - {da_D\over du}{d\ba\over d\bu}\right)\cr
&+ {\cal F}_{\mu\nu}\partial^{\nu}u\left({da\over du}\right)^{-1}
\left({d^2a\over du^2}{d\ba_D\over d\bu} 
- {d^2a_D\over du^2}{d\ba\over d\bu}\right)\cr
&- {\bar{\cal F}}_{\mu\nu}\partial^{\nu}\bu
\left({d\ba\over d\bu}\right)^{-1}
\left({d^2\ba\over d\bu^2}{d\ba_D\over d\bu}-
{d^2\ba_D\over d\bu^2}{d\ba\over d\bu}\right)\ .\cr}
\eqn\bilineartwo
$$
Recalling  that $\tau = da_D/da$ one easily obtains the equation
of motion
$$
0= \partial^{\nu}{\cal F}_{\mu\nu}(\tau-{\bar \tau})
+ {\cal F}_{\mu\nu}\partial^{\nu}u{d\tau\over du}
- {\bar{\cal F}}_{\mu\nu}\partial^{\nu}\bu{d{\bar \tau}\over d\bu} \ .
\eqn\eqofmvect
$$
Examining the real and imaginary
parts of \eqofmvect\ we find
$$\eqalign{
0 &= \partial_{[\lambda} F_{\mu\nu]}\ ,\cr
0 &= {\rm Im}(\partial_{\mu}(\tau{\cal F}^{\mu\nu}))\ . \cr}
\eqn\reim
$$
Thus the choice of $F$ given in \H\ does indeed
obey the standard Bianchi identity, justifying our
ansatz for $H$.
These are precisely the vector equation of motion obtained 
from the Seiberg-Witten effective action. 

While the above derivation of the vector equation of motion using differential
forms is simple and
direct, the analogous procedure for the scalar equation of motion is not
so straightforward.
To this end let us consider another derivation of \eqofmvect . Now we simply
start from
$$\eqalign{
0 &= \int_{\Sigma} \star dH\wedge\lambdabar \cr
&= \int_{\Sigma} E_{\mu z}dz\wedge \lambdabar \ ,\cr}
\eqn\surface
$$
and directly substitute the same expressions in for $H$. In principle
one may consider any
one form in  the integral in 
\surface, rather than $\lambda$. 
However, since one needs the integrand to be well defined over
the entire Riemann surface, there is effectively a unique choice, i.e.
the holomorphic one form. Thus we find 
$$\eqalign{
0 = &\partial^{\nu}{\cal F}_{\mu\nu}\left({da\over du}\right)^{-1}I_0
-{\cal F}_{\mu\nu}\partial^{\nu}u{d^2 a\over du^2}
\left({da\over du}\right)^{-2}I_0\
+ {\cal F}_{\mu\nu}\partial^{\nu}u\left({da\over du}\right)^{-1}{dI_0\over du}
\cr&
-{\cal F}_{\mu\nu}\partial^{\nu}u \left({da\over du}\right)^{-1}J
+{\bar{\cal F}}_{\mu\nu}\partial^{\nu}\bu 
\left({d\ba\over d\bu}\right)^{-1}K
\ ,\cr}
\eqn\surfacetwo
$$
where
$$\eqalign{
I_0 &= \int_{\Sigma}\lambda\wedge\lambdabar \ ,\cr
J&= R^2\Lambda^4\int_{\Sigma}\partial_z\left(
{\lambda_z^2\partial_{\bz}\bs\over 1+R^2\Lambda^4\partial_z s\partial_{\bz}\bs}
\right)dz\wedge\lambdabar\ ,\cr
K &= R^2\Lambda^4 \int_{\Sigma}\partial_{z}\left(
{\lambdabar_{\bz}^2\partial_{z}s\over 1+R^2\Lambda^4\partial_z s\partial_{\bz}
\bs}
\right)dz\wedge\lambdabar \ .\cr}
\eqn\Idefs
$$
Here we see that we arrive at some non-holomorphic integrals over $\Sigma$.
While it is straightforward to evaluate $I_0$ using the Riemann Bilinear
relation to find
$$
I_0 = {da_D\over du}{d\ba\over d\bu} - {da\over du}{d\ba_D\over d\bu} \ ,
\eqn\Io
$$
the $J$ and $K$ integrals require a more sophisticated analysis 
to evaluate them directly. However,
upon comparing \surfacetwo\ with \bilineartwo\ we learn that 
$$\eqalign{
J &= 0\ , \cr
K &= \left({d^2\ba\over d\bu^2}{d\ba_D\over d\bu}-
{d^2\ba_D\over d\bu^2}{d\ba\over d\bu}\right)\ ,\cr
&= -\left({d\ba\over d\bu}\right)^{2}{d{\bar \tau}\over d\bu}\ .\cr}
\eqn\JK
$$
Only with these identifications to we find that the  equation of
motion for the vectors is obtained in agreement with the first method. We
will see that the $J$ and $K$ integrals will appear again in the scalar
equation. We will discuss the explicit evaluation of these integrals elsewhere.


\chapter{The Scalar Equation of Motion}

In this section we will derive the equation of motion for the scalar
zero modes when the vectors are non-zero. As seen in equation \Escalar\ 
above  the
equation of motion for the scalar zero modes in six dimensions is just 
$E=0$.
To reduce this equation to four dimensions  we consider the analogue of
\surface\
$$
0 = \int_{\Sigma} Edz\wedge\lambdabar \ .
\eqn\scalartwo
$$
If we note that 
$\partial_{\mu}s = \lambda_{z}\partial_{\mu}u$ and substitute 
\H\ for the three form we find{
$$
0 = \partial^{\mu}\partial_{\mu}u I_0 
+ \partial_{\mu}u\partial^{\mu}u {dI_0\over du} 
- \partial_{\mu}u\partial^{\mu} u J
- 16{\bar {\cal F}}_{\mu\nu}{\bar {\cal F}}^{\mu\nu}
\left({d\ba\over d\bu}\right)^{-2}K .
\eqn\scalartwo
$$
Thus again the $I_0$, $J$ and $K$ integrals appear. Since we have deduced
the values of these integrals previously we may simply write down the
four-dimensional scalar equation of motion as
$$\eqalign{
0=&\partial^{\mu}\partial_{\mu}u
{da\over du}{d\ba\over d\bu}(\tau-{\bar \tau}) + 
\partial^{\mu}u\partial_{\mu}u
\left({da\over du}{d\ba\over d\bu}{d\tau\over du}
+(\tau-{\bar \tau}){d^2a\over du^2}{d\ba\over d\bu}\right) \cr
&+ 16{\bar {\cal F}}_{\mu\nu}{\bar {\cal F}}^{\mu\nu}
{d {\bar \tau}\over d\bu}\ .\cr}
\eqn\scalarthree
$$
This may then be rewritten as
$$
0=\partial^{\mu}\partial_{\mu}a(\tau-{\bar \tau}) + 
\partial^{\mu}a\partial_{\mu}a{d\tau\over da} 
+ 16 {\bar {\cal F}}_{\mu\nu}{\bar {\cal F}}^{\mu\nu}
{d {\bar \tau}\over d\ba} \ .
\eqn\eqofmscalar
$$
Thus we have obtained the complete low energy effective equations of
motion for  the 
M-fivebrane in the presence of threebranes. We note that both
\eqofmvect\ and \eqofmscalar\ can be obtained from the four-dimensional action
$$
S_{SW} = \int d^4 x\ {\rm Im} \left(  
\tau\partial_{\mu}a\partial^{\mu}{\bar a}
+16 \tau{\cal F}_{\mu\nu}{\cal F}^{\mu\nu}\right)\ ,
\eqn\action
$$
which
is precisely the bosonic part of 
the full Seiberg-Witten effective action for 
$N=2$ supersymmetric $SU(2)$ Yang-Mills [\SW].


\chapter{Higher Derivative Terms}

In the absence of the vectors the dynamics for the scalars 
can be encoded in the more familiar $p$-brane action  
$$
S_5= M_p^6\int d^6 x \sqrt {-{\rm det} g_{\hat m \hat n}}\ ,
\eqn\scalaraction 
$$
where $M_p$ is the eleven-dimensional Planck mass.
It was shown in [\HLW] that the terms only quadratic in spacetime derivatives  
were precisely those of the Seiberg-Witten action. In [\HLW] it was  
pointed out that \scalaraction\ predicts an infinite number of
higher derivatives terms  and the fourth order correction was explicitly  
given.
In this section we examine these terms in more detail. 
The determinant of $g_{\hat n\hat m}$ can be shown to be given by  
$$
\sqrt{-\det g_{\hat n\hat m}}=\left(1+{1\over2}\partial_{\hat m}s\partial^{
\hat m}\bar s\right)\left\{1-{1\over4}{|\partial_{\hat m}s\partial^{\hat m}s|^2
\over(1+{1\over2}
\partial_{\hat m}s\partial^{\hat m}\bar s)}\right\}^{1\over2}\ .
\eqn\detexp
$$
It is straightforward to substitute this expression into the above action 
and then Taylor expand in the number of spacetime derivatives. The result 
after subtracting a total derivative is 
$$\eqalign{ 
S_5=
& {M^6_pR^2\over \Lambda^4}{1\over 2i}
\int d^4xd z\wedge d {\bar z}\left\{
{1\over 2}\partial_{\mu} u\partial^{\mu}{\bar u}{1\over Q{\bar Q}}\right.\cr
&\left. + \sum^{\infty}_{n=1, p=0}R^{2p+4n-2}
C_{n,p}{(\partial_\mu u\partial^\mu\bar u)^p|\partial_\mu u\partial^\mu u|^{2n}
\over (Q\bar Q)(Q\bar Q+4R^2\Lambda^4 z \bar z)^{p+2n-1}}\right\} \ ,\cr}
\eqn\expansion
$$
where 
$$
C_{n,p}=(-1)^n\left({1\over2}\right)^{2n+p}\left({{1\over 2}\atop n}\right)
\left({-2n+1\atop p}\right), \ \ n\geq 1,\quad p\ge 0\ ,
\eqn\Cnpdef
$$ 
and $\left(n \atop m\right)$ is the $m$th binomial coefficient in the
expansion of $(1+x)^n$. 
The first term is none other than the Seiberg-Witten action. Rewriting 
the remaning terms, \expansion\ can be cast in the form 
$$\eqalign{
S_5= &{M^6_pR^2\over \Lambda^4}{1\over 2i} \int d^4x\left\{ {1\over 2}
\partial_\mu u \partial^\mu {\bar u} I_0\right.\cr
&\left.+ \sum^{\infty}_{n=1, p=0} {C_{n,p}I_{(p+2n-1)}}
(\partial_\mu u\partial^\mu\bar u)^p
|\partial_\mu u\partial^\mu u|^{2n} \right\}\ ,\cr}
\eqn\expansiontwo
$$
where 
$$\eqalign{
I_{k}&= {1\over \Lambda^{4k}}\int dz\wedge d{\bar z}
\left({1\over Q \bar Q}\right)
\left({R^2\Lambda^4\over Q\bar Q+4R^2\Lambda^4  z {\bar z}}\right)^{k}\ ,\cr
&={1\over \Lambda^{6k+2}}\int d z_0 \wedge d{\bar z}_0
\left({1\over Q_0 \bar Q_0}\right)
\left({\sigma^2\over Q_0{\bar Q}_0+4\sigma^2 z_0 {\bar z}_0}\right)^{k} 
\ .\cr}
\eqn\Ipdef
$$
Here $\sigma =R\Lambda$, $z_0 = z/\Lambda$, 
$Q_0 =\sqrt{(z_0^2 + u_0)^2 - 1}$ and  $u_0=u/\Lambda^2$.
This form for $I_k$ makes it clear that they obey the scaling relation
$I_k(\rho^{-1}R,\rho \Lambda,\rho^2 u) = \rho^{-6k-2} I_k (R, \Lambda, u)$.
The integrals $I_{k}$ that occur in the action are finite, 
that is there are no   singularities of  the integrand on the Riemann 
surface which lead the integrals to diverge. Given the type of 
singularities that can occur at $z=\infty$ and at the roots of $Q$ 
it is remarkable to examine how the above integral avoids  
the possible divergences. This is presumably a tribute to the 
consistency of M theory. 

Using the Cauchy-Schwartz inequality we may place a bound on the integrals 
$$\eqalign{
I_{k} &\le  {1\over \Lambda^{4k}}
\left|\int d z \wedge d\bar z {1\over Q\bar Q}\right|
\left|\int d z \wedge d\bar z 
\left({R^2\Lambda^4\over Q\bar Q+4 R^2\Lambda^2 z\bar z}\right)^k\right| \cr
&\le R^{2k}\left|\int d z \wedge d\bar z 
{1\over Q\bar Q}\right|^{k+1}\ .}
\eqn\ineq
$$
To obtain the last line we used the fact that 
$|Q\bar Q|\le |Q\bar Q+4 R^2\Lambda^4 z\bar z|$. The expression that occurs 
in the final line is  just that for the Seiberg-Witten action 
which we can evaluate exactly. In particular in the region $|u|\to \infty $ 
we therefore find that 
$$
I_{k} \le R^{2k} 
\left(\left| {da \over du}\right|^2{\rm Im}\tau \right)^{k+1}
\approx  R^{2k}\left|{{\rm ln} u \over u}\right|^{k+1}\ .
\eqn\bound
$$

The most interesting question is whether the above higher derivative 
terms which originate from the classical fivebrane equations of motion 
are related to those that occur in the $N=2$ Yang-Mills theory.  
Yang-Mills theory has an effective action which in principle
depends on $g_{YM},\hbar,u$ and the renormalisation scale $\mu$. However,
as can be seen from the classical action,  
$g_{YM}$ and $\hbar$ always appear as $\hbar g^2_{YM}$. It is also well known
from the renormalisation group that $\hbar g^2_{YM}$
and $\mu$ appear as a single scale $\Lambda_{QCD}$ in the quantum theory. 
Thus the low energy
effective action for Yang-Mills theory only
depends on $\Lambda_{QCD}$ and $u$.
By comparing $I_0$ with the the 
Seiberg-Witten solution one learns that $\Lambda = \Lambda_{QCD}$. However
the extra parameter $\sigma$  appears in 
$I_k$ for $k\ge 1$, hence the higher derivative terms 
of \scalaraction\ also depend on $\sigma$.
From this  observation
it is clear that the higher derivative terms in \expansiontwo\ can never
reproduce those obtained from the Yang-Mills equations. The appearance
of the extra parameter in the M-fivebrane effective action reflects
the fact that these are really the long wave length equations of a
self-dual string theory, not a Yang-Mills theory.
However, let us see how they 
qualitatively compare. 
The higher derivative terms of Yang-Mills theory are of the form 
$$
\int d^4 x \int d^4\theta d^4 \bar \theta K(A, \bar A)\ .
\eqn\Kaction
$$
Unfortunately, the exact form of $K$ is not  not known, 
but arguments have been given to suggest that $K$ has the form 
(at lowest order) 
$K\propto {\rm ln} A \ {\rm ln} \bar A$ [\Grisaru].
Using the expression of [\Grisaru] we may evaluate this in terms of $N=1$ 
superfields from which it is apparent that the result is of the form 
(ignoring logarithmic corrections) 
$|\partial_\mu a \partial^\mu \bar a|^2 |a|^{-4}$, 
in the region of large $a$. 

This must be compared with the first term in the expansion \expansion.
Due to its subtle form, it is difficult to evaluate the integral $I_1$, 
even in the large $u$ limit.
One could assume that the dominant contributions come from the
zeroes in $Q_0$ or make the substitution $Q_0=|z_0|^2 + u_0$. 
Both these approximations are consistent with the bound \bound\ and
lead to the behaviour $I_1 \approx k|u|^{-2}$ where the
constant  $k$ is independent
of $\sigma$. Since 
we have in this region  $u\propto a^2$,  we must 
conclude that if the suggested higher derivative corrections to the 
$N=2$ Yang-Mills theory are correct and our approximation methods reliable
then the higher derivative 
corrections obtained from  the classical M-fivebrane dynamics 
have a weaker fall-off for large $u$ than those of the Yang-Mills theory. 
Since it is believed that these additional 
terms come from a string theory it
is natural to see that the high energy behaviour is qualitatively different
from that of a field theory.


\chapter{Discussion}

In this paper we have presented the complete details concerning the
evaluation of  the bosonic
low energy effective action for threebrane solitons in the M-fivebrane.
In particular we explicitly derived the equations of motion for the
vector degrees of freedom and verified that they exactly reproduce
those of the Seiberg-Witten effective action. We also discussed an 
infinite number of higher derivative terms predicted by the M-fivebrane
theory and compared them with the expected higher derivative terms of
$N=2$ $SU(2)$ Yang-Mills. We note here that the 
generalisation to the gauge group $SU(N)$ is
easily obtained by considering $N$ threebranes with $N-1$ moduli $u_i$
as in [\HLW]. In this case the natural choice for the three form is
$H_{\mu\nu z}= (ds/du_i){\cal F}^i_{\mu\nu}$. In addition one
may also consider $SO(N)$ and  $Sp(N)$
groups by substituting in the curves of [\BSTY] for $F(t,s)$.

Finally, since one motivation of this
paper was to set up the appropriate formalism to calculate four-dimensional
effective actions with $N=1$ supersymmetry, let us briefly describe how this
might work. The simplest generalisation would be to consider threebrane
solitons obtained by the intersection of three M-fivebranes over a threebrane.
This is achieved by turning on an addition complex scalar $w = X^7 + iX^8$.
Following [\three] one again finds that a configuration with 
$X^6,X^{10}$ and $X^7,X^{8}$ active will preserve one quarter  of the 
M-fivebrane
supersymmetry, provided that both $s$ and $w$ are holomorphic functions 
of $z$. Furthermore the M-fivebrane equations of motion will also be 
automatically satisfied in this case. Thus, a
configuration with both $s$ and $w$ holomorphic will lead to a
threebrane soliton with $N=1$ supersymmetry on its worldvolume with both
vector and scalar zero modes. Once can then follow the analysis presented in
this paper and derive the low energy effective equations of motion for both
the vector and scalar zero modes. However, unlike the $N=2$ case considered
here, there will be no supersymmetry which relates the two. It would be
interesting to see if the correct quantum corrections to the low energy
effective action are also predicted 
by these models.

\endpage

\noindent Note Added

While this paper was being written we received a copy of [\dHOO] which has some
overlap with section six of this paper. 
There the derivation 
of the Seiberg-Witten effective action from the M-fivebrane presented in  
[\HLW] was repeated. The first higher derivative
correction found in [\HLW] is considered  and it is concluded 
that these terms are not from a Yang-Mills theory.

\refout

\end